\documentstyle[12pt]{article}
\begin{document}
\begin{center}
{\bf Low Energy Tests of Chiral Symmetry\footnote{Invited talk given
at PANIC 96, Williamsburg, VA}}\\
\medskip
Barry R. Holstein\\
Department of Physics and Astronomy\\
University of Massachusetts\\
Amherst, MA  01003\\
\medskip
{\bf Abstract}
\end{center}
\begin{quote}
The present status of low energy tests of chiral invariance via chiral
perturbation theory is reviewed, both in the meson and baryon sectors,
and future prospects are discussed.
\end{quote}
\section{Introduction}
When I was a student back in the 1960's, the goal of particle and
nuclear physicists was to seek the fundamental laws of nature, and one
of the requirements of such a ``fundamental'' 
law was that it be renormalizable.  Well now it's 1996 and we've
learned a few things in these three decades.  One is that
nonrenormalizable effective field theories can be just as if not more
useful than their renormalizable siblings in certain situations.  One
of these is the case of QCD, where we have what we feel is a correct
model of nature.  However, it is written in terms of the ``wrong''
degrees of freedom (quarks and gluons rather than hadrons) and is
impossible to solve because of its strong coupling and inherent
nonlinearity.  Much more useful in the arena of low energy physics is
an effective Lagrangian, which is written in terms of experimental
degrees of freedom---mesons and baryons---and 
which encodes the symmetries of the underlying
QCD interaction---specifically for our purposes chiral symmetry, which
exists in the limit in which the quark mass can be taken as vanishing.
This is a program which was begun in the 60's with the effective
two-derivative Lagrangian\cite{2}
\begin{equation}
{\cal L}^{(2)}={F_\pi^2\over 4}{\rm Tr}D_\mu UD^\mu U^\dagger
+{F_\pi^2\over 4}{\rm Tr}2B_0m(U+U^\dagger)
\end{equation}
which describes the interaction of the Goldstone fields $\phi_i,
i=1..8$.  Here $F_\pi=92.4$ MeV is the pion decay constant, $m$ is the
quark mass matrix, $B_0$ is a phenomenological constant and 
\begin{equation}
U=\exp({i\over F_\pi}\sum_{j=1}^8\lambda_j\phi_j)
\end{equation}
is a nonlinear function of the fields which transforms as
$LUR^\dagger$ under chiral rotations.  When used at tree level this
interaction is rather successful in predicting low energy
interactions.\cite{2}  For example, expanding to fourth order in
the fields one finds the well-known Weinberg predictions for $\pi\pi$
scattering lengths\cite{3} 
\begin{equation}
a_0^0={7m_\pi\over 32\pi F_\pi^2}\qquad a_0^2=-{m_\pi\over 16\pi F_\pi^2}
\end{equation}
which are confirmed by experiment.  However, in
order to go further and include loop effects one must include
additional four-derivative pieces into the effective Lagrangian, with
ten phenomenological constants $L_i, i=1...10$ which can be determined
experimentally as shown by Gasser and Leutwyler,\cite{4} yielding values for
these parameters as given in Table 1.
\begin{table}
\begin{center}
\begin{tabular}{c|c|c}
Coefficient & Value & Origin\\
\hline
$L_1$&$0.65\pm 0.28$&$\pi\pi$ scattering\\
$L_2$&$1.89\pm 0.26$& and\\
$L_3$&$-3.06\pm 0.02$&$K_{\ell 4}$ decay\\
$L_5$&$2.3\pm 0.2$&$F_K/F_\pi$\\
$L_9$&$7.1\pm 0.3$&$<r_\pi^2>$\\
$L_{10}$&$-5.6\pm 0.3$&$\pi\rightarrow e\nu\gamma$
\end{tabular}
\end{center}
\caption{Gasser-Leutwyler coefficients $L_i$ and the means of determination.}
\end{table}
\section{Mesons}
As mentioned above, the one loop chiral expansion has been carried out
in the cae of Goldstone boson interactions by many investigators.  As
emphasized by Weinberg,\cite{5} this is basically an expansion in
energy-momentum with a scale
parameter $\Lambda_\chi\sim 1$ GeV, so that one is 
entitled to quit at one loop provided that energies are small compared
to this scale.  It is this for this reason that this is called chiral
{\it perturbation} theory ($\chi$pt).  Although ten seems at first like a large
number of parameters, $\chi$pt is very predictive and the extent to
which these predictions are valid is at some level a probe of the
validity of QCD itself.  This subject has been extensively reviewed in
many places\cite{7} and there is in general very good agreement
between predicted and measured quantities as shown in Table 2
\begin{table}
\begin{center}
\begin{tabular}{l|l|c|c}
Reaction & Quantity& Theory& Expt.\\
\hline
$\gamma\rightarrow\pi^+\pi^-$&$<r_\pi^2>\,({\rm fm}^2)$& 0.44
(input)&
$0.44\pm 0.02$\\
$\gamma\rightarrow K^+K^-$&$<r_K^2>\,({\rm fm}^2)$&0.44&$0.34\pm 0.05$\\
$\pi^+\rightarrow e^+\nu_e\gamma$& $h_V(m_\pi^{-1})$&0.027&$0.029\pm
0.017$\\
\quad&$h_A/h_V$&0.46\,({\rm input})&$0.46\pm 0.08$\\
$K^+\rightarrow
e^+\nu_e\gamma$&$(h_A+h_V)(m_\pi^{-1})$&0.038&$0.043\pm 0.003$\\
$\pi^+\rightarrow e^+\nu_ee^+e^-$&$r_A/h_V$&2.6&$2.6\pm 0.6$\\
$\gamma\pi^+\rightarrow\gamma\pi^+$&$(\alpha_E+\beta_M)(10^{-4})\,{\rm
fm}^3$&0&$1.4\pm 3.1$\\
\quad&$\alpha_E(10^{-4}{\rm fm}^3)$&2.8&$6.8\pm 1.4$\cite{8}\\
\quad&\quad&\quad& $2.2\pm 1.1$\cite{9}\\
$K\rightarrow\pi e^+\nu_e$&$ \xi=f_-(0)/f_+(0)$&-0.13& $-0.20\pm
0.08$\\
\quad&$\lambda_+\,({\rm fm}^2)$&0.067&$0.065\pm 0.005$\\
\quad&$\lambda_0\,({\rm fm}^2)$&0.040&$0.050\pm 0.012$
\end{tabular}
\end{center}
\caption{Comparison between chiral predictions and experimental values
for parameters in the Goldstone sector.}
\end{table}
The one area here where there is a possible problem has to do with the
required relationship between the charged pion polarizability 
$\alpha_E^\pi$ and the axial structure function $h_A$ in radiative
pion decay\cite{10}
\begin{equation}
\alpha_E^\pi={\alpha h_A\over \sqrt{2}F_\pi m_\pi}
\end{equation}
The chirally required value for the polarizability---$2.8\times
10^{-4}$ fm$^3$---is at variance with the value found at Serpukov via
radiative pion scattering\cite{8} but not with that found at SLAC in
$\gamma\gamma\rightarrow 
\pi^+\pi^-$.\cite{9}  This is clearly an area which
deserves further study and work in this regard is presently underway
at DA$\Phi$NE, Fermilab and Mainz.
\section{Baryons}
In the case of pion-nucleon intereactions, chiral perturbative
calculations can also be performed.  However, things are much less
clean for reasons which will become clear.  One begins as before with
the simplest $\pi N$ Lagrangian having chiral symmetry
\begin{equation}
{\cal L}_{\pi N}=\bar{N}(i\not\!\!{D}-m_N+{g_A\over 2}\not\!{u}\gamma_5)N
\end{equation}
where $D_\mu$ is a covariant derivative, $g_A$ is a coupling
constant to be determined, and $u_\mu=iu^\dagger\nabla_\mu Uu^\dagger$
with $u^2=U$.  Expanding to lowest order we find
\begin{equation}
{\cal L}_{\pi N}=\bar{N}(i\not\!{\partial}-m_N-ig_A\vec{\tau}\cdot
\not\!\!{\vec{A}}\gamma_5-{g_A\over
F_\pi}\vec{\tau}\cdot\not\!{\nabla}
\vec{\phi}+\ldots )N\label{eq:zz}
\end{equation}
so that $g_A$ is to be identified with the axial coupling in neutron
beta decay.  Also we see that chiral symmetry requires the
Goldberger-Treiman
relation between $g_A$ and the $\pi NN$ coupling constant\cite{20}
\begin{equation}
F_\pi g_{\pi NN}=m_Ng_A
\end{equation}
Using the best present values we have
\begin{equation}
1201\, {\rm MeV}=92.4\, {\rm MeV}\times 13.0\quad{\rm vs.}\quad 939\, 
{\rm MeV}\times 1.26=1183\, {\rm MeV}
\end{equation}
and the agreement to better than two percent strongs confirms the
validity of chiral invariance in the nucleon sector.  A second probe
in this arena arises from the feature that the nucleon matrix element
of the axial current also includes a pion pole contribution, leading
to a prediction that in the muon capture process one requires\cite{21}
\begin{equation}
r_P={g_P(q^2=-0.9m_\mu^2)\over g_A(q^2=-0.9m_\mu^2)}=
{2m_N m_\mu\over m_\pi^2+0.9m_\mu^2}=7.0
\end{equation}
which can be experimentally checked.  Present results are
\begin{equation}
r_P=\left\{\begin{array}{cc}
7.4\pm 2.0\cite{22} & \mu^-\,{\rm capture}\, {}^3{\rm He}\\
6.5\pm 2.4\cite{23} & \mu^-\,{\rm capture}\, {\rm H}\\
10\pm 1\cite{24} & {\rm radiative}\,\mu^-\,{\rm capture}\, {\rm H}
\end{array}\right.
\end{equation}
Obviously the discrepancy in the case of the radiative capture needs
to be further explored.

In order to go further one requres loop corrections, just as in the
mesonic case.  A problem arises here that for the nuclons one
has an additional dimensionful paramter---$m_N$---which is the same
size as the chiral scale, which makes the entire renormalization
procedure doubtful.   This problem can be gotten around, however, by
using so-called heavy baryon methods, which are equivalent to the use
of a Foldy-Wouthuysen transformation and which make a consistent power
counting scheme possible.  Of course, renormalization introduces new
low energy constants (six, {\it e.g.} at ${\cal O}(p^2)$) but
nevertheless this program has been carried out, predominantly by the
group of Bernard, Kaiser, and Meissner,\cite{26} and applications are reported
in many systems:
\begin{table}
\begin{center}
\begin{tabular}{cl}
$\pi N\rightarrow\pi N$ & Scattering lengths, $\sigma_{\pi N}$ \\
$\pi N\rightarrow \pi\pi N$ & LET's, $\pi\pi$ Scattering lengths \\
$\gamma N\rightarrow \gamma N$ & Polarizabilities, DHG sum rule \\
$\gamma N\rightarrow \pi N$ & LET's \\
$\gamma^*N\rightarrow  \pi N$& LET's, $g_A(q^2)$ \\
$\gamma N\rightarrow \pi\pi N$ & Chiral loop effects in $\pi^0\pi^0$
\end{tabular}
\end{center}
\caption{Examples of nucleon reactions which have been examined via $\chi$pt.}
\end{table}
One area of particular interest here is that of pion photoproduction.
In this case for charged production the feature that the pion
derivative in Eq. \ref{eq:zz} is covariant leades to the
Kroll-Ruderman predictions for the $E_{0+}$ (electric dipole)
multipole at threshold
\begin{equation}
E_{0+}^{\rm th}=\left\{
\begin{array}{cc}
\sqrt{2}D(1-{3\over 2}\mu+{\cal O}(\mu^2)) & \pi^+n \\
-\sqrt{2}D(1-{1\over 2}\mu+{\cal O}(\mu^2)) & \pi^-p
\end{array}\right.
\end{equation}
where $D=eg_{\pi NN}/8\pi m_N=23.9\,(\times 10^{-3}/m_\pi)$ and
$\mu\equiv m_\pi/m_N$.  In this
case theory and experiment are in good agreement, as shown below 
\begin{table}
\begin{center}
\begin{tabular}{c|c|c}
\quad & theory & expt. \\
\hline
$E_{0+}^{\rm th}(\pi^+n)$ & 26.3 &$ 27.9\pm 0.5\cite{27} $\\
\quad & \quad & $28.8\pm 0.7$\cite{28} \\
$E_{0+}^{\rm th}(\pi^-p)$ & -31.3 & $-31.4\pm 1.3\cite{27}$\\
\quad &\quad & $-31.2\pm 1.2\cite{29}$
\end{tabular}
\end{center}
\caption{Threshold values of $E_{0+}$ for charged pion photoproduction
($\times 10^{-3}/m_\pi$).}
\end{table}
However, the experimental numbers quoted are from old emulsion
measurements and resutls from the recent SAL experiment are eagerly
awaited.

In the case of neutral pion photoproduction, things are more
interesting.  In this case the venerable Low Energy Theorems (LET)
for the threshold $E_{0+}$ multipole
\begin{equation}
E_{0+}^{\rm th}=\left\{
\begin{array}{ll}
-D(\mu-{1\over 2}\mu^2(3+\kappa_p)+{\cal O}(\mu^3))=-2.3 & \pi^0p \\
-D({1\over 2}\mu^2\kappa_n+{\cal O}(\mu^3))=+0.5 & \pi^0n
\end{array}\right.
\end{equation}
were shown in 1991 to be incorrect due inappropriate assymptions
concerning analyticity.\cite{31}  New loop contributions at ${\cal O}(\mu^2)$
\begin{equation}
\Delta E_{0+}^{\rm th}=-D\mu^2({m_N\over 4F_\pi})^2
\end{equation}
are large and destroyed the agreement which appeared to exist between
the original LET predictions and preliminary $\pi^0 p$ experiments at
both Mainz and at Saclay.  However, there has been a good deal of
recent activity.  On the theoretical side, Bernard et al. have
performed an analysis at ${\cal O}(p^4)$.\cite{31}  In doing so they require
the values of two counterterms.  Estimating these via
$\Delta,\rho,\omega$ dominance, they predict a value for $E_{0+}$ in
good agreement with new experiments performed at both MAMI and SAL
\begin{table}
\begin{center}
\begin{tabular}{c|c|c}
\quad & {\rm theory} & {\rm expt.}\\
\hline
$E_{0+}(\pi^0p) (\times 10^{-3}/m_\pi)$ & -1.2 &$ -1.31\pm 0.08$\cite{33} \\
\quad &\quad & $-1.3\pm 0.5\pm 0.6$\cite{34}\\
$P_1/|\vec{q}|(\pi^0p)(\times {\rm GeV}^{-2})$ & 0.480 &$ 0.47\pm 0.01$\cite{33}\\
\quad&\quad& $0.41\pm 0.03\cite{34}$
\end{tabular}
\end{center}
\caption{Threshold parameters for neutral pion photoproduction.}
\end{table}
This may be somewhat accidental, as the convergence of the
series appears slow---$E_{0+}=C(1-1.26+0.59+\ldots)$.  However, it has
been pointed out that the P-wave calculations do not suffer from this
slow convergence, yielding what should be a very solid prediction
\begin{eqnarray}
{1\over |\vec{q}|}P_1&\equiv&(M_{1+}-M_{1-}+3E_{1+})\nonumber\\
&=&{D\over M_N}[1+\kappa_p-\mu
(1+{1\over 2}\kappa_p-{g_{\pi NN}^2(10-3\pi)\over
48\pi})+\ldots]=0.480\, {\rm GeV}^{-2}
\end{eqnarray}
There may be a remaining problem in the size of the $E_{1+}$
multipole,\cite{34} but this involves a significant theoretical cancellation.

A second arena of activity is that of Compton scattering.  In
this case recent precise measurements of both the proton and neutron
polarizabilities have been performed yielding values as shown below.
As can be observed, the electric polarizability of the neutron is
comparable to but slightly larger than its proton counterpart.  This
is interesting since a valence quark model cannot produce such a
result, yielding instead a prediction\cite{40}
\begin{equation}
\alpha_E^p-\alpha_E^n={\alpha\over 3M_N}(<r_p^2>-<r_n^2>)\approx 4.6
(\times 10^{-4}\,{\rm fm}^3)
\end{equation}
On the other hand a chiral expansion of the polarizability does not
have this problem, starting off from equal values for the proton and
neutron\cite{41}
\begin{equation}
\alpha_E^p=\alpha_E^n=10\beta_M^p=10\beta_M^n={e^2g_A^2\over
192\pi^3F_\pi^2M_N}({5\pi\over 2\mu}+\ldots)=12.4 (\times 10^{-4}\,{\rm fm}^3)
\end{equation}
A full ${\cal O}(p^4)$ calculation of both electric and magnetic
polarizabilities, with counterterms evaluated via resonance dominance
yields very satisfactory results as shown Table 6, although
again the convergence of the series is in doubt.\cite{42}  Similarly a good
deal of work has been done in the case of polarized Compton
scattering, both experimentally and theoretically, but we do not have
the space to discuss this here.
\begin{table}
\begin{center}
\begin{tabular}{c|c|c}
\quad & theory & expt.\\
\hline
$\alpha_E^p$& 10.5 & $11.6\pm 0.6\pm 0.6$\cite{43}\\
$\beta_M^p$&3.5 & $2.6\mp 0.6\mp 0.6$\cite{43}\\
$\alpha_E^n$&13.4&$12.6\pm 1.5\pm 2.0$\cite{44}\\
$\beta_M^n$&7.8 & $3.2\mp 1.5\mp 2.0\cite{44}$
\end{tabular}
\end{center}
\caption{Experimental values of electric and magnetic polarizabilities
compared to chiral predictions at ${\cal O}(p^4)$.  (All $\times
10^{-4}\,{\rm fm}^3)$}
\end{table}
\section{Conclusions}
Obviously in a short report such as this the discussion above can
provide only a brief introduction to the multitude of 
work which is presently underway.
Additional areas of activity include
\begin{itemize}
\item [i)] $\gamma p\rightarrow\pi^0\pi^0 p$ for which the significant
near threshold cross section observed via the TAPS group at Mainz is
being confronted with the pion loop corrected amplitude.  The loop
correction is large but it is too early to tell whether it agrees with
the experimental findings.
\item[ii)] $\gamma^*p\rightarrow\pi^0p$ for which $k^2\simeq -0.1$
GeV$^2$ data from both NIKHEF and MAMI seems to contradict at least 
some of the chiral perturbative predictions.  However, this value of
momentum transfer is probably above the range where one can expect
agreement, and we await the lower $k^2$ data to be taken at Mainz.
\item[iii)] $\pi N\rightarrow\pi\pi N$ for which previous analysis in
order to extract the $\pi\pi$ scattering lengths has utilized the
Olsson-Turner parameterization, which is inconsistent with a modern
chiral analysis.
\item[iv)] $K_{\ell 4}$ measurements at DA$\Phi$NE and elsewhere
should be able to resolve the question concerning the use of standard
vs. generalized chiral perturbation theory.
\item[v)] $\vec{\gamma}\vec{p}\rightarrow\gamma p$ measurements at
CEBAF and elsewhere combined with precise resonance photoproduction 
data should shed light on the validity of the DHG sum rule.
\item[vi)] on the theoretical side it is important to include effects
of the $\Delta(1240)$ as a baryonic degree of freedom and not just as
a heavy state which contributes to a counterterm.  Work to this end is
underway and should appear soon.
\end{itemize}
Overall then I hope that I have been able to convey the sense that
the area of low energy tests of the standard model via chiral
perturbation theory is an active and exciting one, with plenty of work
remaining to be done on both the theoretical {\it and} experimental
sides.

{\bf Acknowledgement:}
This research is supported in part by the National Science Foundation.
It is a pleasure to acknowledge the warm hospitality of the Oberlin
College physics department where this work was performed.

\end{document}